\documentclass[%
 reprint,
 amsmath,amssymb,
 aps,
prb,
floatfix,
]{revtex4-1}

\usepackage{graphicx}
\usepackage{dcolumn}
\usepackage{bm}
\usepackage{amsfonts}
\usepackage{mathrsfs}
\usepackage{bbold}
\usepackage{tikz}
\usetikzlibrary{arrows}

\begin{document}

\title{Generation of a superconducting vortex via N\'eel skyrmions}

\author{J. Baumard$^{1,2}$}
\author{J. Cayssol$^1$}%
\author{F.S. Bergeret$^{2,3}$}%
\author{A. Buzdin$^{1,4}$}%
\affiliation{$^1$Univ. Bordeaux, CNRS, LOMA, UMR 5798, F-33405 Talence, France} 
\affiliation{$^2$Donostia International Physics Center (DIPC), Manuel de Lardizabal 5, E-20018 San Sebasti\'an, Spain}
\affiliation{$^3$Centro de F\'isica de Materiales (CFM-MPC), Centro Mixto CSIC-UPV/EHU, Manuel de Lardizabal 4, E-20018 San Sebasti\'an, Spain}
\affiliation{$^4$Sechenov First Moscow State Medical University, Moscow, 119991, Russia}

\date{\today}

\begin{abstract}
We consider a type-II superconducting thin film in contact with a N\'eel skyrmion. The skyrmion induces spontaneous currents in the superconducting layer, which under the right condition generate a superconducting vortex in the absence of an external magnetic field. We compute the magnetic field and current distributions in the superconducting layer in the presence of N\'eel skyrmion.
\end{abstract}

\maketitle

\section{Introduction}
Superconductor-ferromagnet heterostructures \cite{buzdin_proximity_2005,bergeret_odd_2005,linder_superconducting_2015,blamire_interface_2014} in the presence of spin-orbit and exchange interactions are attracting great interest due to the possible realization of topological qubits based on Majorana fermions \cite{kitaev_unpaired_2001,nayak_non-abelian_2008,oreg_helical_2010,lutchyn_majorana_2010,wu_majorana_2017,black-schaffer_majorana_2011,alicea_new_2012} and the fact that such systems display unconventional magnetoelectric effects \cite{buzdin_direct_2008,*konschelle_magnetic_2009,ojanen_magnetoelectric_2012,pershoguba_currents_2015,chudnovsky_manipulating_2017,konschelle_theory_2015,*bergeret_theory_2015,fominov_josephson_2007,kulagina_spin_2014,konschelle_ballistic_2016, halterman_induced_2008,braude_fully_2007,mironov_spontaneous_2017,silaev_anomalous_2017,robinson_controlled_2010}.  In particular, the interplay between  spin-orbit coupling  and a homogeneous Zeeman or exchange field may lead to spontaneous supercurrents in bulk superconductors and hybrid structures.
From a SU(2) covariant formulation of spin dependent fields, a spin-orbit coupling and homogeneous Zeeman field is equivalent to an inhomogeneous magnetic texture that, in combination with superconducting correlations, may support spontaneous currents under certain symmetry conditions \cite{bergeret_spin-orbit_2014,konschelle_theory_2015,melnikov_interference_2012}.

Among inhomogeneous magnetic textures, skyrmions \cite{bogdanov_thermodynamicallystable_1989,rossler_chiral_2011,leonov_properties_2016} have attracted interest because of their nanoscale dimension (1nm  - 100 nm), topological robustness, and the low current density needed to move them, which makes them good candidates as information carriers in future memory devices \cite{jonietz_spin_2010,yu_skyrmion_2012, kiselev_chiral_2011,iwasaki_current-induced_2013,hrabec_current-induced_2017}. It has been shown that a skyrmion can be stabilized when proximity-coupled to an s-wave superconductor \cite{fraerman_magnetization_2005,vadimov_magnetic_2018}. In addition, such systems can induce sponstaneous currents \cite{rabinovich_chirality_2018}, Majorana bound states \cite{yang_majorana_2016,gungordu_stabilization_2018}, Weyl points \cite{takashima_supercurrent-induced_2016} or Yu-Shiba-Rusinov-like states \cite{pershoguba_skyrmion-induced_2016}. Moreover, \citeauthor{hals_composite_2016} \cite{hals_composite_2016} studied the interaction between a skyrmion and a vortex by assuming that they are stabilized in the magnetic and superconducting layers. 

In this article, we investigate the formation of a composite topological excitation between a magnetic skyrmion and a superconducting vortex in a ferromagnet (F)/superconductor (S) bilayer with Rashba spin-orbit coupling. In contrast to Ref. \onlinecite{hals_composite_2016}, the superconducting vortex is initially absent. We show that the generation of a vortex is via the magnetoelectric effect induced by the skyrmion in the presence of a sufficiently strong spin-orbit coupling. By evaluating the free energy of the F/S system, we derive the conditions required for the creation of this vortex, and compute  the current and magnetic field distributions in the superconductor.\\
The paper is organized as follows. In Sec. \ref{Section: free energy}, we introduce the free energy describing the system. In Sec. \ref{Section: nucleation condition}, we derive the vortex nucleation condition. The magnetic field and current distributions are provided in Sec. \ref{Section: field and current}. We finally conclude and give some perspectives implied by our work in Sec. \ref{Section: conclusion}.

\section{Setup and free energy \label{Section: free energy}}
\begin{figure}
\begin{center}
\includegraphics[scale=0.25]{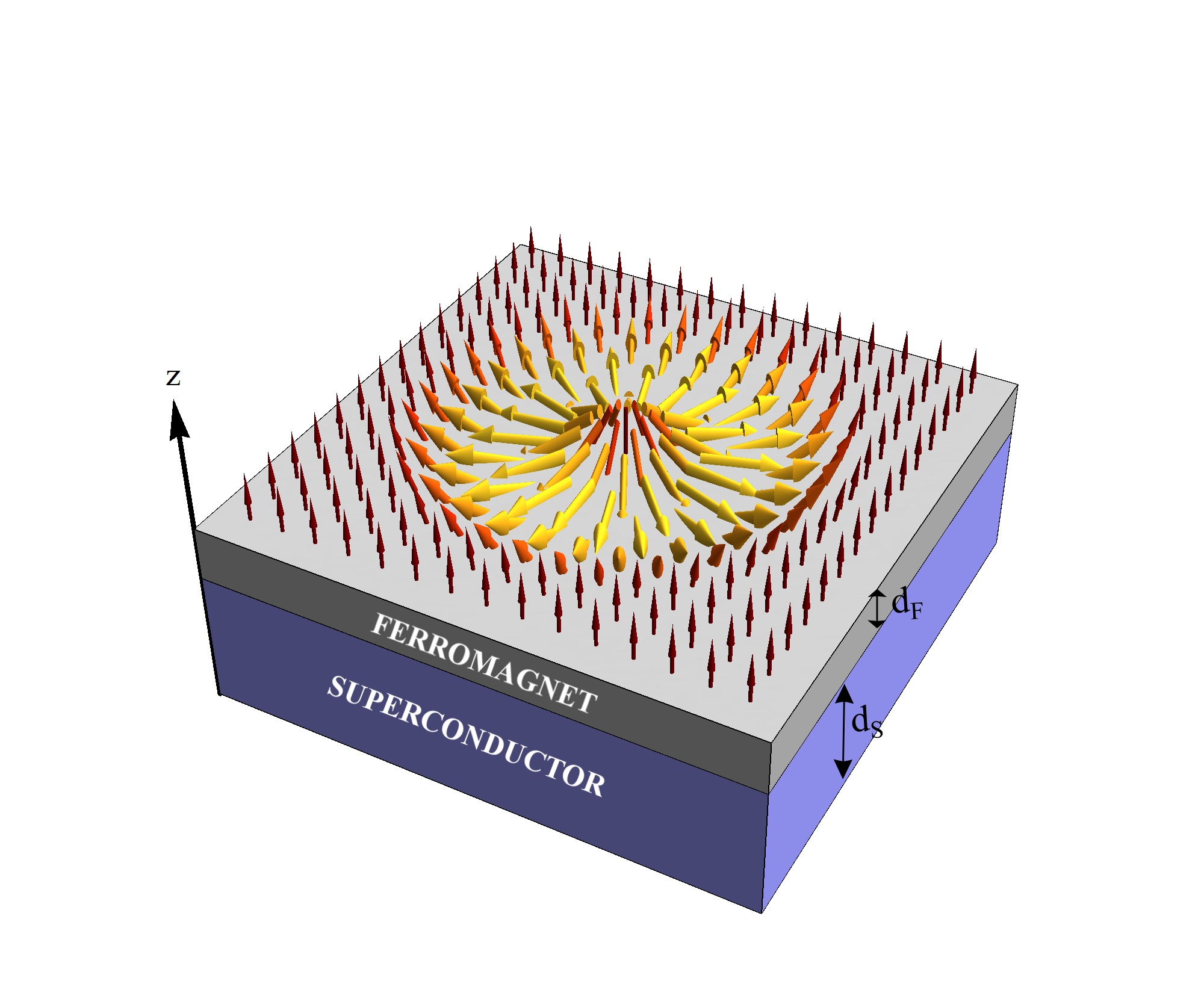}
\end{center}
\caption{A thin superconducting film proximity-coupled to a ferromagnetic layer hosting a N\'eel skyrmion. The F layer has thickness $d_\text{F}$ and the S layer $d_\text{S}$. \label{system_schema}}
\end{figure}

We consider a type-II superconducting thin film of thickness $d_\text{S}$, characterized by the coherence length $\xi$ and the London penetration length $\lambda$. The superconductor is in contact with a ferromagnet of thickness $d_\text{F}$ hosting a N\'eel skyrmion (Fig. \ref{system_schema}). We assume that a  two-dimensional spin-orbit interaction  is present in the ferromagnetic layer and described by the Rashba constant $\alpha_\text{R}$. The N\'eel skyrmion is characterized by the following spin profile \cite{nagaosa_topological_2013} 
\begin{equation}
\vec{S}(\vec{r}) = \eta\,\sin\Theta(r)\,\vec{e}_r + \cos\Theta(r)\,\vec{e}_z ,
\label{eq:S}
\end{equation}
where $\vec{e}_r$ is the radial unit vector and $\vec{e}_z$ the unit vector normal to the F and S layers. The profile function $\Theta(r)$ must obey the boundary conditions $\Theta(0) = \pi$ and $\Theta(\infty) = 0$. For the analytical calculations below, we assume that  $\displaystyle \Theta(r) = \pi\left(1 -r/R\right)$ for $r < R$, and otherwise $0$.  Here $R$ denotes the radius of the skyrmion. The constant $\eta = \pm 1$ describes the skyrmion winding.   The sign of $\eta$, combined with the Rashba constant $\alpha_\text{R}$ determines the vortex polarity. In the following we consider $\alpha_\text{R} > 0$ and $\eta = -1$.

In principle both the direct electromagnetic coupling between the skyrmion and the superconductor \cite{lyuksyutov_ferromagnet_2005}, and the magnetic proximity effect may result in the nucleation of a vortex. In this letter we only focus on the proximity effect by assuming that the exchange field and spin-orbit interaction penetrate the superconductor over the atomic thickness $a$, where $a \ll d_\text{S}$. For a uniform ferromagnetic layer, if the magnetization is smaller than the first critical field, $\mu_0\,M \ll H_{c1}\,$, the standard electromagnetic interaction cannot nucleate a vortex. Even if $M$ exceeds $H_{c1}$, it is possible to avoid vortex formation by designing the F and S layers such that $\mu_0\,M \ll H_{c1}\,\displaystyle\frac{d_\text{S}}{d_\text{F}}$, which may be easily fulfilled if the ferromagnetic layer is much thinner than the superconductor one. 

Let us consider temperatures for which the superconductivity is well developed, \textit{ie.} $T \ll T_c$ . The free energy of the F/S bilayer can be written as
\begin{equation}
F = F_0 + F_{\text{sc}} + F_{\text{L}} + F_{\text{mag}}\; ,
\label{Fgen}
\end{equation}
where $F_0$ is the free energy in the absence of superconductivity and magnetic texture, and 
$F_{\text{sc}}$ is the kinetic term related to the superconducting current energy. To derive the expression of the free energy, we use the London approach which assumes that the generated current does not modify the modulus of the superconducting order parameter. The criterion of applicability of the London approach is well known (see for example Ref. \onlinecite{gennes_superconductivity_1966}): the current density should be much smaller than the critical current density $j_c \propto \displaystyle \frac{\Phi_0}{\mu_0\,\lambda^2\,\xi}$, where $\displaystyle \Phi_0 = \frac{h}{2\,e}$ (with $e >0$) is the superconducting quantum of flux. This is always the case for Abrikosov vortices, except the narrow core region. The computation of the current (see Sec. \ref{Subsection: current}) shows that this approach is completely justified to describe the vortex generation by the skyrmion while $R \gg \xi$. Moreover, since we  assume that $d_\text{S}$ is smaller than $\lambda$,  the density of superconducting current energy is nearly constant through the width $d_\text{S}$ so $F_{\text{sc}}$ reads:
\begin{equation}
F_{\text{sc}} = \int \frac{1}{2\,\mu_0\,\lambda_{\text{eff}}}\left(\vec{\phi}(\vec{r}) - \vec{A}(\vec{r})\right)^2\mathrm{d}^2\vec{r} \, ,
\label{F_kin}
\end{equation}
where $\displaystyle\lambda_{\text{eff}} = \lambda^2/d_\text{S}$ is the effective screening length for the superconductor, $\vec{\phi}$ is the gradient of the local superconducting phase (multiplied by $\hbar/2e$), and $\vec{A}$ is the vector potential. Detailed calculations are provided in Appendix \ref{Appendix:Obtention_FL}. In the presence of a vortex, the expression for  the vector $\vec{\phi}$ can be obtained from the London equation as \cite{gennes_superconductivity_1966}:
\begin{equation}
\vec{\phi}(\vec{r}) = \frac{\Phi_0}{2\,\pi\,r}\vec{e}_{\theta} \, ,
\end{equation}
where $\vec{e}_{\theta}$ is the unit orthoradial vector.

The third contribution to the free energy, $F_{\text{L}}$ in  Eq. (\ref{Fgen}),  corresponds to the coupling energy between the superconductor and the magnetic order induced by the skyrmion.  By proximity effect, the interplay between the exchange field and the Rashba spin-orbit interaction in the ferromagnetic layer induces a spin polarization in the superconducting film. This may give rise for example to a spontaneous current in the bulk superconductor near the interface to F, in the absence of an external magnetic field \cite{edelstein_magnetoelectric_1995,mironov_spontaneous_2017}. For $T$ close to $T_c$, such an interaction is described by the Lifshitz invariant \cite{edelstein_ginzburg_1996,samokhin_magnetic_2004,kaur_helical_2005}.  
At low temperatures and for $d_\text{S} \ll \lambda$, one can consider that the spin-orbit interaction is averaged over $d_\text{S}$. In this case the energy $F_\text{L}$ can be written as:  
\begin{equation}
F_{\text{L}} = \int \vec{\alpha}(r)\cdot\left(\vec{\phi}(\vec{r}) - \vec{A}(\vec{r})\right) \mathrm{d}^2\vec{r} \;,
\label{F_magneto}
\end{equation}
where $\vec{\alpha}(\vec{r}) = \alpha(r)\,\vec{e}_\theta = -\alpha_0\,\sin\Theta(r)\,\vec{e}_\theta$ (see Appendix \ref{Appendix:Obtention_FL}). The constant $\alpha_0$ incorporates the Rashba constant $\alpha_R$, the exchange energy $h_\text{ex}$,  the thickness of the superconducting film $d_\text{S}$ and the proximity length  $a$:
\begin{equation}
\alpha_0 \approx \frac{1}{4\,\mu_0\,e\,\lambda_{\text{eff}}}\,\frac{a}{d_\text{S}}\,\frac{\alpha_R\,h_\text{ex}}{v_F^2} \;.
\end{equation}

The last component $F_{\text{mag}}$ of the free energy Eq. (\ref{Fgen}): 
\begin{equation}
F_{\text{mag}} = \int \frac{\vec{B}^2(\vec{r})}{2\,\mu_0}\mathrm{d}^3\vec{r} \;,
\label{F_mag}
\end{equation}
represents the energy of the magnetic field $\vec{B} = \vec{\nabla}\times\vec{A}$.

The current density for $d_\text{S} \ll \lambda$ in the plane $z = 0$ is given by $\displaystyle \vec{j} = -\frac{\partial f}{\partial \vec{A}}\,\delta(z)$, where $f$ is the free energy density in the film: $F = \int f\, \mathrm{d}^2\vec{r}$. From the Maxwell-Ampere equation $\displaystyle \vec{\nabla}\times \vec{B} = \mu_0\,\vec{j}$, we obtain a differential equation for $\vec{A}$, which can be solved in Fourier space \cite{gennes_superconductivity_1966}. The solution $\vec{A}_q$ of this equation, where $\vec{A}_q$ is the two-dimensional Fourier transform of $\vec{A}$ in the layer, is given by:
\begin{equation}
\vec{A}_q = \frac{1}{1 + 2\,q\,\lambda_{\text{eff}}}\,\left(\vec{\phi}_q + \mu_0\,\lambda_{\text{eff}}\,\vec{\alpha}_q\right) \;,
\label{Aq}
\end{equation} 
where $\vec{\phi}_q$, $\vec{\alpha}_q$ are the two-dimensional Fourier transforms of $\vec{\phi}(\vec{r})$ and $\vec{\alpha}(\vec{r})$ respectively. This calculation is provided in Appendix \ref{Appendix:obtention_A}.

\section{Creation of a superconducting vortex \label{Section: nucleation condition}}
\subsection{Magnetic field induced by the skyrmion}
\begin{figure}
\includegraphics[scale=0.35]{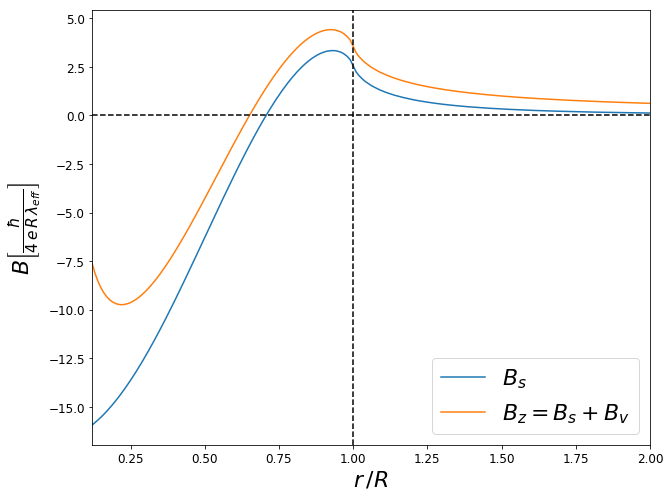}
\caption{Magnetic field distribution in the superconducting layer. The vortex nucleation condition for a vortex with vorticity 1 is fulfilled: $R = 50\,\xi$, $\alpha_R = 0.1\,v_F$ and $h_{\text{eff}} = 20\,k_B\,T_c$. The blue line corresponds to the magnetic field distribution without vortex, whereas the orange one is in the presence of the vortex. \label{champ}}
\end{figure}

Because of the spontaneous current generated by the skyrmion in the superconducting film, a magnetic field is created perpendicular to the layer. We first consider that there is no vortex. In this case, the term proportional to $\vec{\phi}_q$ in Eq.(\ref{Aq}) disappears and we can derive the expression of the magnetic field distribution $\vec{B}_{\text{s}}(r) = B_{\text{s}}(r)\,\vec{e}_z$ in the superconducting layer. Considering that the skyrmion is small compared to $\lambda_{\text{eff}}$ and focusing  on small distances $r$ from the center of the skyrmion ($r \ll \lambda_{\text{eff}}$) one can write;
\begin{equation}
B_{\text{s}}(r) =  -\frac{1}{2}\,\mu_0\,\alpha_0 \int q\,\Gamma(q)\,J_0(q\,r)\,\mathrm{d}q \;, \label{Bs}
\end{equation}
where $\Gamma(q) = \displaystyle \int_0^R r\,\sin(\pi\,\frac{r}{R})\,J_1(q\,r)\,\mathrm{d}r$ and $J_0(q\,r)$, $J_1(q\,r)$ are Bessel functions of first kind. This field distribution is represented by the blue line in Fig. \ref{champ}. As expected, outside of the skyrmion $B_{\text{S}}$ decreases and vanishes very fast. Moreover, one can check that the magnetic flux associated to $B_{\text{s}}$ equals to zero.

\subsection{Vortex nucleation condition}
The condition for the superconducting vortex creation can be derived by comparing the free energy of the system with and without a vortex. We replace $\vec{A}$ by its expression, (Eq.\ref{Aq}), into the contributions  Eq.(\ref{F_kin}, \ref{F_magneto},\ref{F_mag}) to the free energy . The resulting $F$  can be written as a sum of three terms (see Appendix \ref{Appendix:free energy}): 
\begin{equation}
F = F_{\text{v}} + F_{\text{s}} + F_{\text{int}} \;.
\label{F_VS}
\end{equation}
The first one, proportional to $\Phi_0^2$, describes the self-energy of the vortex. The second term, proportional to $\alpha_R^2$, describes the energy of the current induced by the skyrmion, whereas third term, proportional, to $\Phi_0\,\alpha_R$, the interaction energy between the vortex and such  current.

Since $\xi \ll r \ll \lambda_{\text{eff}}$, we can write the self-energy of the vortex in the following way:
\begin{equation}
F_{\text{v}} = \frac{1}{\pi\,\mu_0\,\lambda_{\text{eff}}}\left(\frac{\Phi_0}{2}\right)^2\ln\left(2\,\frac{\lambda_{\text{eff}}}{\xi}\right) \;. \label{Fv}
\end{equation}
By assuming $R \ll \lambda_{\text{eff}}$, we can write the current energy and the interaction term as
\begin{eqnarray}
F_{\text{s}} &=& - \frac{\mu_0}{8\,\pi}\int \alpha_q^2\,\mathrm{d}q \;; \label{F_S}\\
F_{\text{int}} &=&  -\frac{\Phi_0}{2\,\pi\,\mu_0\,e\,\lambda_{\text{eff}}}\frac{a}{d_\text{S}}\frac{\alpha_R\,h_\text{ex}}{v_F^2}\,R \;. \label{Fint}
\end{eqnarray}

The difference of free energy, $\Delta F = F - F_{\text{s}}$, between the states with and without vortex reads: 
\begin{multline}
\Delta F = \frac{\Phi_0^2}{2\,\pi^2\,\mu_0\,\lambda_{\text{eff}}}\left[\frac{\pi}{2}\ln\left(2\,\frac{\lambda_{\text{eff}}}{\xi}\right)\right. \\ \left.-\, 0.180\,\frac{h_\text{ex}}{k_\text{B}\,T_c}\,\frac{a}{d_\text{S}}\,\frac{\alpha_R}{v_\text{F}}\,\frac{R}{\xi}\right] \;,
\label{Delta F}
\end{multline}
where $k_\text{B}$ is the Boltzmann constant, $T_c$ is the critical temperature of the superconductor and 
the coherence length is given by $\xi = \displaystyle 0.180\,\frac{\hbar\,v_F}{k_\text{B}\,T_c}$.

The condition for the vortex nucleation is determined by the condition $\Delta F < 0$, which  gives:
\begin{equation}
\frac{h_{\text{eff}}}{k_\text{B}\,T_c}\,\frac{\alpha_R}{v_F}\,\frac{R}{\xi}> \frac{\pi}{0.36}\,\ln\left(2\,\frac{\lambda_{\text{eff}}}{\xi}\right) \;,
\label{condition}
\end{equation}
where $h_{\text{eff}}=h_\text{ex}\,\displaystyle\frac{a}{d_\text{S}}$ is the average effective exchange energy, with $a \ll d$. \\
The condition Eq.(\ref{condition}) gives the features of the ferromagnetic layer required to induce a vortex inside the superconducting film without any external magnetic field. Qualitatively, this result shows that if $\alpha_R$, $h_{\text{eff}}$ or $R$ increase, so does the magnetic field $B_{\text{s}}$ (Eq. \ref{Bs}), thereby favoring the appearance of the vortex.

\subsection{Multiquanta vortices}
We now discuss the possibility of nucleating a vortex carrying $n$ superconducting flux quanta  $\Phi_0=h/2e$, with $n>1$. So far we only considered the case $n=1$. Let $F_n$ be the free energy in presence of a $n$-quanta vortex.
\begin{equation}
F_n = n^2\,F_{\text{v}} + F_{\text{s}} + n\,F_{\text{int}} \;.
\end{equation}
The optimal value of $n$ can be estimated by minimizing $F_n$ with respect to $n$:
\begin{equation}
n_\text{op} \approx -\frac{F_{\text{int}}}{2\,F_v} = \frac{\pi}{0.72\,\ln(2\frac{\lambda_{\text{eff}}}{\xi})}\frac{h_{\text{eff}}}{k_B\,T_c}\,\frac{\alpha_R}{v_F}\,\frac{R}{\xi} \;.
\end{equation}
Therefore, upon raising the Rashba coupling and/or the exchange field, it is possible to stabilize a multiquanta vortex carrying the integer value of $n_\text{op}$ superconducting flux quanta. However for simplicity  in what follows   we assume that the spin-orbit interaction is  weak enough to have a vortex with vorticity larger than 1.

\section{Magnetic field and current distributions \label{Section: field and current}}
\subsection{Magnetic field}
The presence of the vortex modifies the magnetic field distribution. In addition to the component $B_{\text{s}}$, stemming from the current induced by the skyrmion in the superconducting layer, there is a term originated from the vortex itself. The total magnetic field distribution can thus be written as
\begin{equation}
B_z(r) = B_{\text{s}}(r) + B_{\text{v}}(r) \;. \label{Bz}
\end{equation} 
The term $B_{\text{v}}(r)$ is obtained from the first term of Eq.\ref{Aq}:
\begin{equation}
B_{\text{v}}(r) = \frac{\Phi_0}{4\,\pi\,\lambda_{\text{eff}}\,r} \;, \label{Bv}
\end{equation} 
for $\xi \ll r \ll \lambda_{\text{eff}}$.\\
The magnetic field distribution $B_z(r)$ is shown in Fig. \ref{champ} (orange line). It is assumed that the condition for appearance of a vortex is fulfilled.  As expected, both $B_z$ and $B_{\text{s}}$  follow the spin direction of the skyrmion, with a sinusoidal-like shape: it is negative near the center, and positive for $r \gtrsim 0.65\,R$. At $r = R$, the  amplitude of the magnetic field decreases away from the skyrmion. The component $B_{\text{s}}$ tends to zero very fast, whereas $B_z$ vanishes far from the center. It decreases slowly because of the presence of the vortex, whose component $B_{\text{v}}$ is proportional to $1/r$.

\begin{figure}
\flushleft{a.}
\\
\includegraphics[scale=0.44]{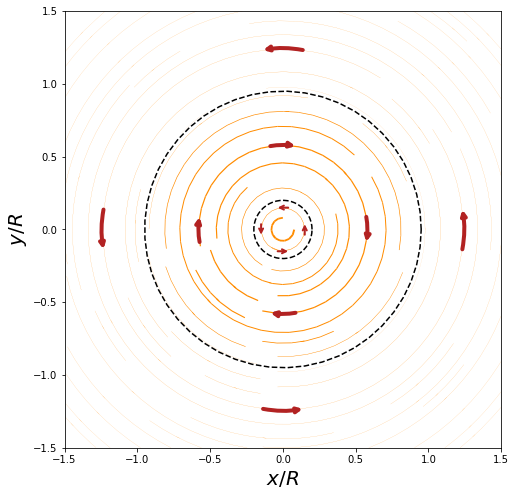}
\\
\flushleft{b.}\\
\includegraphics[scale=0.35]{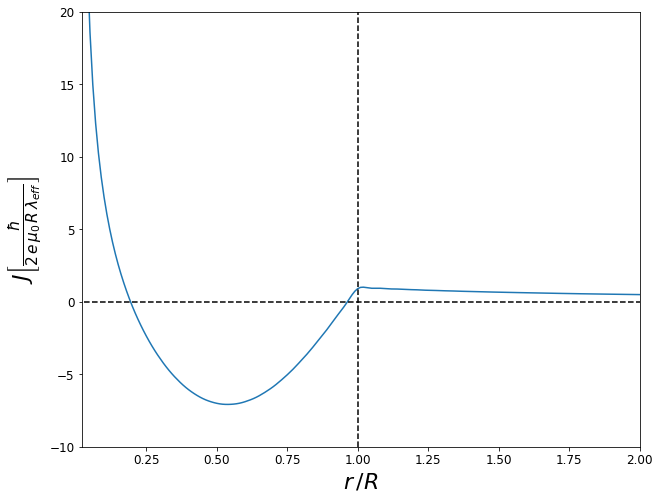}
\caption{a. Current lines in the superconducting layer. The vortex nucleation condition for a vortex with vorticity 1 is fulfilled: $R = 50\,\xi$, $\alpha_R = 0.1\,v_F$ and $h_{\text{eff}} = 20\,k_B\,T_c$. The dashed black lines represent the changes in the rotation direction of the current loops. The thickness of the lines represents the amplitude of the current. b. Distribution of the current $J$ in the superconducting layer.}
\label{fig_courant}
\end{figure}

\subsection{Charge Current \label{Subsection: current}}
The current $\vec{J}$ in the superconducting layer is obtained  from $\vec{J} = -\displaystyle \frac{\partial f}{\partial \vec{A}}$ (see Appendix \ref{Appendix: current}). As with the magnetic field distribution, it can be written as the sum of two contributions: one induced directly by the skyrmion, and a second stemming from the vortex.
\begin{equation}
\vec{J}(r) = \vec{J}_{\text{s}}(r) + \vec{J}_{\text{v}}(r) \; \label{J(r)}.
\end{equation}
Under the same assumptions as before,  $\xi \ll r \ll \lambda_{\text{eff}}$ and $R \ll \lambda_{\text{eff}}$, one obtains
\begin{eqnarray}
\vec{J}_{\text{s}}(r) &=&  -\alpha_0 \int q\,\Gamma(q)\,J_1(q\,r)\,\mathrm{d}q \, \vec{e}_{\theta} \;;\\
\vec{J}_{\text{v}}(r) &=& \frac{\Phi_0}{2\,\pi\,\mu_0\,\lambda_{\text{eff}}\,r}\,\vec{e}_{\theta} \;.
\end{eqnarray}
The current lines in the film are shown  in Fig. \ref{fig_courant}.a. We use the same parameters as in  (Fig. \ref{champ}). 

Around the vortex ($r < 0.20\, R$), the current is dominated by the contribution from the vortex and  flows anticlockwise. As it can be seen Fig. \ref{fig_courant}.b,  in this region the current is positive and decreases like $1/r$.   
For larger values of $r$ within the skyrmion ($0.20\,R < r < 0.95\, R$), the current distribution has a sinusoidal shape, and is dominated by the contribution from the skyrmion. In this region, the current loops are clockwise. Finally, for $r > 0.95\, R$, the current is again dominated by the contribution from the vortex. It decreases slowly with distance, and tends to zero far from the skyrmion.

\section{Conclusion \label{Section: conclusion}}
We have shown that the spontaneous current induced by the skyrmion in the superconducting thin film gives rise to a magnetic field perpendicular to the layer. If the Rashba coupling exceeds a threshold value (given by Eq. \ref{condition}), the skyrmion can nucleate a superconducting vortex by magnetoelectric proximity effect in the absence of an applied external field. The vorticity is determined by the sign of the spin-orbit interaction and the skyrmion winding. Finally we outline some perspectives implied by our calculations. Even if the Rashba coupling threshold condition is not reached, it is possible to nucleate vortices in S simply by applying an external magnetic field (larger than $H_{c1}$), or in the case when vortices are created directly via the electromagnetic coupling with the ferromagnetic layer. In the former situation (external magnetic field), our free energy calculations demonstrate an attractive coupling which will pin vortices to the skyrmion for one orientation of the magnetic field. For the opposite orientation, the vortices should be pushed away by the skyrmion. Such decoration/antidecoration of the skyrmion by vortices can be, in principle, detected  experimentally. Finally, we also stress that the inverse effect, namely the nucleation of a skyrmion via the proximity of a superconducting vortex, is also suggested by our results, a strong effect that could in principle be observed experimentally via magnetic force microscopy or topological Hall effect in systems like Nb/Co/Pt \cite{hrabec_measuring_2014}. \\

\section*{Acknowledgements}
The authors gratefully acknowledge J. W. A. Robinson for his useful remarks and suggestions. This work was supported by EU Network COST CA16218 (NANOCOHYBRI) and the French ANR project SUPERTRONICS and OPTOFLUXONICS (A. B. and J. C.). J. B. and F.S.B. acknowledge funding by the Spanish Ministerio de Econom\'ia y Competitividad (MINECO) (Projects No. FIS2014-55987-P and No. FIS2017-82804-P). J. B. aknowledges the financial support from the Initiative d'Excellence (IDEX) of the Universit\'e de Bordeaux.

\appendix
\section{Derivation of the magnetoelectric energy $\mathbf{F_{\text{L}}}$ \label{Appendix:Obtention_FL}}
We derive the expression of the coupling energy between the superconductor and magnetic order induced by the skyrmion. We start from the Ginzburg-Landau free energy 
\begin{equation}
    F^{\text{GL}} = F^{\text{GL}}_0 + \int \frac{1}{4\,m}\left|\hat{D}\Psi\right|^2\mathrm{d}^3\vec{r} + F^{\text{GL}}_{\text{L}} \;, \label{GL}
\end{equation}
 where $\hat{D} = \left(-\text{i}\,\hbar\,\vec{\nabla} + 2\,e\,\vec{A}\right)$ is the gauge-invariant momentum operator and $\Psi = \left|\Psi\right|e^{\text{i}\,\vec{q}\cdot\vec{r}}$. The term $F_0^{\text{GL}}$ contains all the terms without derivative of $\Psi$. The Lifshitz invariant $F^{\text{GL}}_{\text{L}}$ reads \cite{edelstein_ginzburg_1996,samokhin_magnetic_2004,kaur_helical_2005}:
 \begin{equation}
     F^{\text{GL}}_{\text{L}} = \int\varepsilon(r) \left(\vec{e}_z\times\vec{S}\right)\cdot\left[\Psi^{\star}\,\hat{D}\,\Psi + \text{h.c.}\right]\,\mathrm{d}^3\vec{r}\;.
\end{equation}
One can derive an estimate of $\varepsilon (r)$, which is constant in the region where the spin-orbit interaction is present and null elsewhere. \\
To obtain the expression of the wave-vector $\vec{q}$, we must minimize $F^{\text{GL}}$ with respect to $\vec{q}$ for $\vec{A} = \vec{0}$. We get:
\begin{equation}
    q = -\frac{4\,m}{\hbar}\varepsilon \;.\label{q_calculated}
\end{equation}
From Ref. \onlinecite{dimitrova_theory_2007}, we have an estimate of $q$:
\begin{equation}
    q = \frac{\alpha_R\,h_\text{ex}}{\hbar\,v_F^2}\;. \label{q_estimated}
\end{equation}
By comparing the expressions \ref{q_calculated} and \ref{q_estimated}, one obtains an estimate of $\varepsilon$:
\begin{equation}
    \varepsilon = -\frac{\alpha_R\,h_\text{ex}}{4\,m\,v_F^2}\;.
\end{equation}
For $T \ll T_c$, superconductivity is well developed. We then  rewrite the free energy \ref{GL} in the London approach by noticing that $\Psi = \left|\Psi\right|e^{\text{i}\,\varphi}$, where $\left|\Psi\right|$ is constant and such that $\left|\Psi\right|^2 = \frac{1}{2}\,n_s$ where $n_s$ is the density of superconducting electrons. Thus, the free energy becomes: 
\begin{multline}
    F = F^{\text{L}}_0 + \int \frac{e^2\,n_s}{2\,m}\left(\vec{\phi} - \vec{A}\right)^2\mathrm{d}^3\vec{r}\\
    - e\,n_s\,\varepsilon\int \left(\vec{e}_z\times\vec{S}\right)\cdot\left(\vec{\phi} - \vec{A}\right)\,\mathrm{d}^3\vec{r}\;, \label{F_L}
\end{multline}
where $\vec{\phi} = -\displaystyle\frac{\Phi_0}{2\,\pi}\,\vec{\nabla}\varphi$ and $F^{\text{L}}_0 = F_0 + F_{\text{mag}}$. In what follows, $F_0$ will be omitted.\\
We introduce the London coherence length: 
\begin{equation}
    \lambda^2 = \frac{m}{\mu_0\,n_s\,e^2}\;.
\end{equation}
Considering that $d_\text{S} \ll \lambda$, the quantity $\vec{\phi} - \vec{A}$ is almost constant over $d_\text{S}$. We emphasize that the spin-orbit interaction and the exchange field penetrate the superconducting layer over a distance $a$, corresponding to the atomic thickness. We also assume that the magnetization in the ferromagnetic layer is weak, thus the Zeeman field is negligible compared to the exchange field. Then we can compute the integrals of Eq. \ref{F_L} over the z-direction:
\begin{multline}
    F = F_{\text{mag}} + \frac{d_\text{S}}{2\,\mu_0\,\lambda^2}\int \left(\vec{\phi} - \vec{A}\right)^2\mathrm{d}^2\vec{r}\\
    - e\,n_s\,\varepsilon\,a\int \left(\vec{e}_z\times\vec{S}\right)\cdot\left(\vec{\phi} - \vec{A}\right)\,\mathrm{d}^2\vec{r}\;. \label{F_L integrated}
\end{multline}
Introducing the effective screening length $\lambda_{\text{eff}}$, one can notice that the second term of the free energy \ref{F_L integrated} is exactly the superconducting current energy $F_{\text{sc}}$ (Eq. \ref{F_kin}), whereas the third term corresponds to the magnetoelectric energy $F_{\text{L}}$ (Eq. \ref{F_magneto}).

\section{Final expression of the free energy $\mathbf{F}$}
\subsection{Derivation of the vector potential $\vec{\mathbf{A}}$ \label{Appendix:obtention_A}}
\begin{figure}
\begin{center}
\begin{tikzpicture}[scale=1.5]
	\draw[->, cyan, thick] (0,0) -- (1.5,0);
	\draw (1.5, 0) node[right, cyan]{$\vec{r}$} ;
	\draw[->, blue, thick](0, 0)--(0.5, 0);
	\draw(0.5, 0) node [below, blue]{$\vec{e}_r$};
	\draw[->, green, thick] (0,0) -- (1, 1);
	\draw (1, 1) node[above right, green]{$\vec{q}$} ;
	\draw (1,0) arc(0:45:1);
	\draw(1,0.4) node[right]{$\theta$};
	\draw[->, gray, thick](0, 0)--(0.4, 0.4);
	\draw(0.4, 0.4) node [below right, gray]{$\vec{e}_q$};
	\draw[->] (0,0) -- (0, 1.5);
	\draw[->, blue, thick](0, 0)--(0, 0.5);
	\draw(0, 0.5) node [right, blue]{$\vec{e}_{\theta}$};
	\draw[->, dashed] (0,0) -- (-1, 1);
	\draw[->, gray, thick](0, 0)--(-0.4, 0.4);
	\draw(-0.4, 0.4) node [below left, gray]{$\vec{e}_\perp$};
\end{tikzpicture}
\caption{The sets of coordinates $\left(\vec{e}_r,\, \vec{e}_{\theta}\right)$ and $\left(\vec{e}_q,\, \vec{e}_{\perp}\right)$, respectively corresponding to the real space and the Fourier space. \label{Fig: Unit vectors}}
\end{center}
\end{figure}
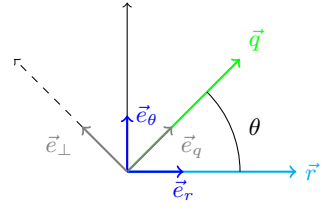

In this section, we derive the expression of the vector potential $\vec{A}$. Let $f$ be the free energy density per unit surface.
\begin{equation}
    f = \frac{1}{2\,\mu_0\,\lambda_{\text{eff}}}\left(\vec{\phi} - \vec{A}\right)^2 + \vec{\alpha}(r)\cdot\left(\vec{\phi} - \vec{A}\right) + \int \frac{B^2}{2\,\mu_0}\mathrm{d}z \;. \label{free energy density}
\end{equation}
The current density for $d_\text{S} \ll \lambda$ in the plane $z = 0$ is given by 
\begin{equation}
\vec{j} = -\frac{\partial f}{\partial \vec{A}}\,\delta(z) = \frac{1}{\mu_0\,\lambda_{\text{eff}}}\left(\vec{\phi} - \vec{A}\right)\,\delta(z) + \vec{\alpha}(r)\,\delta(z) \;. \label{current_density}
\end{equation}
We recall the Maxwell-Ampere equation in the London gauge:
\begin{equation}
    \mu_0\,\vec{j} = \vec{\nabla}\times\vec{B} = -\Delta\vec{A} \;. \label{M-A equation}
\end{equation}
By replacing the expression of the current density \ref{current_density} into the Maxwell-Ampere equation (Eq. \ref{M-A equation}), one gets:
\begin{equation}
    -\Delta\vec{A} + \frac{1}{\lambda_{\text{eff}}}\,\vec{A}\,\delta(z) = \frac{1}{\lambda_{\text{eff}}}\,\vec{\phi}\,\delta(z) + \mu_0\,\vec{\alpha}(r)\,\delta(z) \;. \label{Eq. A}
\end{equation}
We introduce the following three and two-dimensional Fourier transforms:
\begin{eqnarray}
    \vec{A}_{qk} &=& \int \vec{A}(\vec{r},z)\,e^{\text{i}\left(\vec{q}\cdot\vec{r} + k\,z\right)}\mathrm{d}^2\vec{r}\,\mathrm{d}z \;;\label{Aqk}\\
    \vec{A}_q &=& \frac{1}{2\,\pi}\int \vec{A}_{qk}\mathrm{d}k = \int \vec{A}(\vec{r})\,e^{\text{i}\,\vec{q}\cdot\vec{r}}\mathrm{d}^2\vec{r} \;;\label{A_q} \\
    \vec{\phi}_q &=& \int \vec{\phi}(r)\,e^{\text{i}\,\vec{q}\cdot\vec{r}}\mathrm{d}^2\vec{r} =  \text{i}\,\frac{\Phi_0}{q}\,\vec{e}_\perp \;;\label{phi_q}\\
    \vec{\alpha}_q &=& \int \vec{\alpha}(r)\,e^{\text{i}\,\vec{q}\cdot\vec{r}}\mathrm{d}^2\vec{r} = \text{i}\,\alpha_q\,\vec{e}_{\perp} \;, \label{alpha_q}
\end{eqnarray}
with $\alpha_q = 2\,\pi \displaystyle \int_0^{\infty}r\,\alpha(r)\,J_1(q\,r)\,\mathrm{d}r$ where $J_1(q\,r)$ is a Bessel function of first kind. The unit vector $\vec{e}_\perp$ is represented in Fig. \ref{Fig: Unit vectors}.\\

By taking the Fourier transform of Eq. \ref{Eq. A}, one obtains the following  equation:
\begin{equation}
    \vec{A}_{qk} = \frac{1}{q^2 + k^2}\left[\frac{1}{\lambda_{\text{eff}}}\left(\vec{\phi}_q - \vec{A}_q\right) + \mu_0\vec{\alpha}_q\right] \;. \label{Eq. Aqk}
\end{equation}
After integration of Eq. \ref{Eq. Aqk} over $k$, one recovers the expression of $\vec{A}_q$ (Eq. \ref{Aq}).

\subsection{Free energy $\mathbf{F}$ \label{Appendix:free energy}}
In this section, we explain how we obtain the expression \ref{F_VS} for the free energy $F$ and how we determine the value of $\eta$ depending on the sign of $\alpha_\text{R}$.\\
Each term of $F$ (Eq. \ref{Fgen}) can be written in terms of $\vec{A}_q$, $\vec{\phi}_q$ and $\vec{\alpha}_q$:
\begin{eqnarray}
	F_\text{sc} &=& \frac{1}{\left(2\,\pi\right)^2}\,\frac{1}{2\,\mu_0\,\lambda_\text{eff}} \int \left|\vec{\phi}_q - \vec{A}_q\right|^2\,\mathrm{d}^2\vec{q} \;; \label{Fsc-q}\\
    F_\text{L} &=& \frac{1}{\left(2\,\pi\right)^2} \int \vec{\alpha}_q^{\star}\cdot\left(\vec{\phi}_q - \vec{A}_q\right)\mathrm{d}^2\vec{q} \;; \label{Fex-q}\\
    F_\text{mag} &=& \frac{1}{\left(2\,\pi\right)^3} \int \frac{\left|\vec{B}_{qk}\right|^2}{2\,\mu_0}\,\mathrm{d}^2\vec{q}\,\mathrm{d}k \;, \label{Fmag-q}
\end{eqnarray}
where $\vec{B}_{qk}$ is the Fourier transform of $\vec{B}$, and is such that
\begin{equation}
	\left|\vec{B}_{qk}\right|^2 = k^2\left|\vec{A}_{qk}\right|^2 + \left|\left(\vec{q}\times\vec{A}_{qk}\right)\cdot\vec{e}_z\right|^2 \;. \label{Bqk}
\end{equation}
We replace $\vec{A}_q$ (Eq. \ref{A_q}) and $\vec{A}_{qk}$ (Eq. \ref{Aqk}) by their expression in Eq. \ref{Fsc-q} to \ref{Fmag-q}. After integration, one obtains:
\begin{align}
F_\text{sc} &= \frac{\lambda_\text{eff}}{4\,\pi\,\mu_0} \int \frac{q}{\left(1 + 2\,q\,\lambda_\text{eff}\right)^2}\left(2\,\Phi_0 - \mu_0\,\alpha_q\right)^2\mathrm{d}q \;; \label{Fsc-q 2}\\
F_\text{L} &= \frac{1}{2\,\pi} \int \frac{\lambda_\text{eff}\,q}{1 + 2\,q\,\lambda_\text{eff}}\left(2\,\Phi_0 - \mu_0\,\alpha_q\right)\alpha_q\mathrm{d}q \;; \label{Fex-q 2}\\
F_\text{mag} &= \frac{1}{2\,\pi\,\mu_0} \int \frac{\left(\Phi_0 + \mu_0\,\lambda_\text{eff}\,q\,\alpha_q\right)^2}{\left(1 + 2\,q\,\lambda_\text{eff}\right)^2}\mathrm{d}q \;. \label{Fmag-q 2}
\end{align}
Using Eq. \ref{Fsc-q 2} to \ref{Fmag-q 2}, one can rewrite $F$ as a sum of three terms. The first one, proportional to $\Phi_0^2$, is called $F_\text{v}$ (Eq. \ref{Fv}). The second term is proportional to $\alpha_q^2$, and corresponds to $F_\text{s}$ (Eq. \ref{F_S}),  and the third one, depending on the product $\Phi_0\,\alpha_q$, is called $F_\text{int}$ (Eq. \ref{Fint}).\\
In order to nucleate a vortex in the superconducting layer, the difference of energy $\Delta F = F - F_{\text{s}}$ must be negative:
\begin{multline}
\Delta F = \frac{\Phi_0^2}{2\,\pi^2\,\mu_0\,\lambda_{\text{eff}}}\left[\frac{\pi}{2}\ln\left(2\,\frac{\lambda_{\text{eff}}}{\xi}\right)\right. \\ \left.+\, 0.180\,\eta\,\frac{h_\text{ex}}{k_\text{B}\,T_c}\,\frac{a}{d_\text{S}}\,\frac{\alpha_R}{v_\text{F}}\,\frac{R}{\xi}\right] \;.
\label{Delta F}
\end{multline}
The condition $\Delta F < 0$ requires that $\eta\,\alpha_\text{R} < 0$: the polarity of the vortex is determined by the spin-orbit interaction and the skyrmion winding. We thus consider $\alpha_\text{R} > 0$, which implies $\eta = -1$.

\section{Magnetic field and current distributions \label{Appendix: field and current}}
\subsection{Perpendicular magnetic field distribution $\mathbf{B(r)}$ \label{Appendix:magnetic field}}
We compute the normal component $B_z(r)$ of the magnetic field distribution, which is given by $B_z(r) = \left(\vec{\nabla}\times\vec{A}\right)\cdot\vec{e}_z$. In the Fourier space, this relation becomes
\begin{equation}
    B^z_{qk} = -\text{i}\left(\vec{q}\times\vec{A}_{qk}\right)\cdot\vec{e}_z \;, \label{Eq. Bqk}
\end{equation}
where $\vec{A}_{qk}$ is obtained from Eq. \ref{Eq. Aqk} after replacing $\vec{A}_q$ by its expression (Eq. \ref{Aq}). Thus
\begin{equation}
    B^z_{qk} = \frac{1}{q^2 + k^2}\,\frac{2\,q}{1 + 2\,q\,\lambda_{\text{eff}}}\left(\Phi_0 + q\,\lambda_{\text{eff}}\,\mu_0\,\alpha_q\right) \;. \label{Bqk^z}
\end{equation}
After integration over $k$, we get the component $B^z_q$, which is the Fourier transform of $B_z(r)$:
\begin{equation}
    B^z_q = \frac{1}{1 + 2\,q\,\lambda_{\text{eff}}}\left(\Phi_0 + q\,\lambda_{\text{eff}}\,\mu_0\,\alpha_q\right) \;. \label{Bq}
\end{equation}
After taking the inverse Fourier transform of Eq. \ref{Bq}, in the approximation $q \gg \lambda_{\text{eff}}^{-1}$ one obtains the expression \ref{Bz}. Notice that Eq. \ref{Bv} was previously obtained in Ref. \onlinecite{gennes_superconductivity_1966}.

\subsection{Current in the superconducting layer \label{Appendix: current}}
In this section, we derive the expression of the current $\vec{J}(r)$ in the superconducting layer, obtained from the free energy density (Eq. \ref{free energy density}):
\begin{equation}
    \vec{J}(r) = -\frac{\partial f}{\partial \vec{A}} = \frac{1}{\mu_0\,\lambda_{\text{eff}}}\left(\vec{\phi} - \vec{A}\right) + \vec{\alpha}(r) \;.
\end{equation}
In the Fourier space and after replacing $\vec{A}_q$ by its expression (Eq. \ref{Aq}), the current becomes:
\begin{equation}
    \vec{J}_q = \frac{2\,\text{i}}{1 + 2\,q\,\lambda_{\text{eff}}}\left(\frac{\Phi_0}{\mu_0} + q\,\lambda_{\text{eff}}\,\alpha_q\right)\vec{e}_\perp \;. \label{Jq}
\end{equation}
Taking into account that $\vec{e}_\perp = -\sin\theta\,\vec{e}_r + \cos\theta\,\vec{e}_\theta$ (see Fig. \ref{Fig: Unit vectors}), one can perform the inverse Fourier transform of \ref{Jq}. In the approximation $q \gg \lambda_{\text{eff}}^{-1}$, we finally obtain Eq. \ref{J(r)}.

\bibstyle{apsrev4-1}
\bibliography{References}

\begin{thebibliography}{53}%
\makeatletter
\providecommand \@ifxundefined [1]{%
 \@ifx{#1\undefined}
}%
\providecommand \@ifnum [1]{%
 \ifnum #1\expandafter \@firstoftwo
 \else \expandafter \@secondoftwo
 \fi
}%
\providecommand \@ifx [1]{%
 \ifx #1\expandafter \@firstoftwo
 \else \expandafter \@secondoftwo
 \fi
}%
\providecommand \natexlab [1]{#1}%
\providecommand \enquote  [1]{``#1''}%
\providecommand \bibnamefont  [1]{#1}%
\providecommand \bibfnamefont [1]{#1}%
\providecommand \citenamefont [1]{#1}%
\providecommand \href@noop [0]{\@secondoftwo}%
\providecommand \href [0]{\begingroup \@sanitize@url \@href}%
\providecommand \@href[1]{\@@startlink{#1}\@@href}%
\providecommand \@@href[1]{\endgroup#1\@@endlink}%
\providecommand \@sanitize@url [0]{\catcode `\\12\catcode `\$12\catcode
  `\&12\catcode `\#12\catcode `\^12\catcode `\_12\catcode `\%12\relax}%
\providecommand \@@startlink[1]{}%
\providecommand \@@endlink[0]{}%
\providecommand \url  [0]{\begingroup\@sanitize@url \@url }%
\providecommand \@url [1]{\endgroup\@href {#1}{\urlprefix }}%
\providecommand \urlprefix  [0]{URL }%
\providecommand \Eprint [0]{\href }%
\providecommand \doibase [0]{http://dx.doi.org/}%
\providecommand \selectlanguage [0]{\@gobble}%
\providecommand \bibinfo  [0]{\@secondoftwo}%
\providecommand \bibfield  [0]{\@secondoftwo}%
\providecommand \translation [1]{[#1]}%
\providecommand \BibitemOpen [0]{}%
\providecommand \bibitemStop [0]{}%
\providecommand \bibitemNoStop [0]{.\EOS\space}%
\providecommand \EOS [0]{\spacefactor3000\relax}%
\providecommand \BibitemShut  [1]{\csname bibitem#1\endcsname}%
\let\auto@bib@innerbib\@empty
\bibitem [{\citenamefont {Buzdin}(2005)}]{buzdin_proximity_2005}%
  \BibitemOpen
  \bibfield  {author} {\bibinfo {author} {\bibfnamefont {A.~I.}\ \bibnamefont
  {Buzdin}},\ }\href {\doibase 10.1103/RevModPhys.77.935} {\bibfield  {journal}
  {\bibinfo  {journal} {Rev. Mod. Phys.}\ }\textbf {\bibinfo {volume} {77}},\
  \bibinfo {pages} {935} (\bibinfo {year} {2005})}\BibitemShut {NoStop}%
\bibitem [{\citenamefont {Bergeret}\ \emph {et~al.}(2005)\citenamefont
  {Bergeret}, \citenamefont {Volkov},\ and\ \citenamefont
  {Efetov}}]{bergeret_odd_2005}%
  \BibitemOpen
  \bibfield  {author} {\bibinfo {author} {\bibfnamefont {F.~S.}\ \bibnamefont
  {Bergeret}}, \bibinfo {author} {\bibfnamefont {A.~F.}\ \bibnamefont
  {Volkov}}, \ and\ \bibinfo {author} {\bibfnamefont {K.~B.}\ \bibnamefont
  {Efetov}},\ }\href {\doibase 10.1103/RevModPhys.77.1321} {\bibfield
  {journal} {\bibinfo  {journal} {Rev. Mod. Phys.}\ }\textbf {\bibinfo {volume}
  {77}},\ \bibinfo {pages} {1321} (\bibinfo {year} {2005})}\BibitemShut
  {NoStop}%
\bibitem [{\citenamefont {Linder}\ and\ \citenamefont
  {Robinson}(2015)}]{linder_superconducting_2015}%
  \BibitemOpen
  \bibfield  {author} {\bibinfo {author} {\bibfnamefont {J.}~\bibnamefont
  {Linder}}\ and\ \bibinfo {author} {\bibfnamefont {J.~W.~A.}\ \bibnamefont
  {Robinson}},\ }\href {\doibase 10.1038/nphys3242} {\bibfield  {journal}
  {\bibinfo  {journal} {Nat. Phys.}\ }\textbf {\bibinfo {volume} {11}},\
  \bibinfo {pages} {307} (\bibinfo {year} {2015})}\BibitemShut {NoStop}%
\bibitem [{\citenamefont {Blamire}\ and\ \citenamefont
  {Robinson}(2014)}]{blamire_interface_2014}%
  \BibitemOpen
  \bibfield  {author} {\bibinfo {author} {\bibfnamefont {M.~G.}\ \bibnamefont
  {Blamire}}\ and\ \bibinfo {author} {\bibfnamefont {J.~W.~A.}\ \bibnamefont
  {Robinson}},\ }\href {\doibase 10.1088/0953-8984/26/45/453201} {\bibfield
  {journal} {\bibinfo  {journal} {J. Phys. Condens. Matter}\ }\textbf {\bibinfo
  {volume} {26}},\ \bibinfo {pages} {453201} (\bibinfo {year}
  {2014})}\BibitemShut {NoStop}%
\bibitem [{\citenamefont {Kitaev}(2001)}]{kitaev_unpaired_2001}%
  \BibitemOpen
  \bibfield  {author} {\bibinfo {author} {\bibfnamefont {A.~Y.}\ \bibnamefont
  {Kitaev}},\ }\href {\doibase 10.1070/1063-7869/44/10S/S29} {\bibfield
  {journal} {\bibinfo  {journal} {Physics-Uspekhi}\ }\textbf {\bibinfo {volume}
  {44}},\ \bibinfo {pages} {131} (\bibinfo {year} {2001})}\BibitemShut
  {NoStop}%
\bibitem [{\citenamefont {Nayak}\ \emph {et~al.}(2008)\citenamefont {Nayak},
  \citenamefont {Simon}, \citenamefont {Stern}, \citenamefont {Freedman},\ and\
  \citenamefont {Das~Sarma}}]{nayak_non-abelian_2008}%
  \BibitemOpen
  \bibfield  {author} {\bibinfo {author} {\bibfnamefont {C.}~\bibnamefont
  {Nayak}}, \bibinfo {author} {\bibfnamefont {S.~H.}\ \bibnamefont {Simon}},
  \bibinfo {author} {\bibfnamefont {A.}~\bibnamefont {Stern}}, \bibinfo
  {author} {\bibfnamefont {M.}~\bibnamefont {Freedman}}, \ and\ \bibinfo
  {author} {\bibfnamefont {S.}~\bibnamefont {Das~Sarma}},\ }\href {\doibase
  10.1103/RevModPhys.80.1083} {\bibfield  {journal} {\bibinfo  {journal}
  {Reviews of Modern Physics}\ }\textbf {\bibinfo {volume} {80}},\ \bibinfo
  {pages} {1083} (\bibinfo {year} {2008})}\BibitemShut {NoStop}%
\bibitem [{\citenamefont {Oreg}\ \emph {et~al.}(2010)\citenamefont {Oreg},
  \citenamefont {Refael},\ and\ \citenamefont {von Oppen}}]{oreg_helical_2010}%
  \BibitemOpen
  \bibfield  {author} {\bibinfo {author} {\bibfnamefont {Y.}~\bibnamefont
  {Oreg}}, \bibinfo {author} {\bibfnamefont {G.}~\bibnamefont {Refael}}, \ and\
  \bibinfo {author} {\bibfnamefont {F.}~\bibnamefont {von Oppen}},\ }\href
  {\doibase 10.1103/PhysRevLett.105.177002} {\bibfield  {journal} {\bibinfo
  {journal} {Phys. Rev. Lett.}\ }\textbf {\bibinfo {volume} {105}},\ \bibinfo
  {pages} {177002} (\bibinfo {year} {2010})}\BibitemShut {NoStop}%
\bibitem [{\citenamefont {Lutchyn}\ \emph {et~al.}(2010)\citenamefont
  {Lutchyn}, \citenamefont {Sau},\ and\ \citenamefont
  {Das~Sarma}}]{lutchyn_majorana_2010}%
  \BibitemOpen
  \bibfield  {author} {\bibinfo {author} {\bibfnamefont {R.~M.}\ \bibnamefont
  {Lutchyn}}, \bibinfo {author} {\bibfnamefont {J.~D.}\ \bibnamefont {Sau}}, \
  and\ \bibinfo {author} {\bibfnamefont {S.}~\bibnamefont {Das~Sarma}},\ }\href
  {\doibase 10.1103/PhysRevLett.105.077001} {\bibfield  {journal} {\bibinfo
  {journal} {Phys. Rev. Lett.}\ }\textbf {\bibinfo {volume} {105}},\ \bibinfo
  {pages} {077001} (\bibinfo {year} {2010})}\BibitemShut {NoStop}%
\bibitem [{\citenamefont {Wu}\ \emph {et~al.}(2017)\citenamefont {Wu},
  \citenamefont {Anderson}, \citenamefont {Hsiao},\ and\ \citenamefont
  {Levin}}]{wu_majorana_2017}%
  \BibitemOpen
  \bibfield  {author} {\bibinfo {author} {\bibfnamefont {C.-T.}\ \bibnamefont
  {Wu}}, \bibinfo {author} {\bibfnamefont {B.~M.}\ \bibnamefont {Anderson}},
  \bibinfo {author} {\bibfnamefont {W.-H.}\ \bibnamefont {Hsiao}}, \ and\
  \bibinfo {author} {\bibfnamefont {K.}~\bibnamefont {Levin}},\ }\href
  {\doibase 10.1103/PhysRevB.95.014519} {\bibfield  {journal} {\bibinfo
  {journal} {Phys. Rev. B}\ }\textbf {\bibinfo {volume} {95}},\ \bibinfo
  {pages} {014519} (\bibinfo {year} {2017})}\BibitemShut {NoStop}%
\bibitem [{\citenamefont {Black-Schaffer}\ and\ \citenamefont
  {Linder}(2011)}]{black-schaffer_majorana_2011}%
  \BibitemOpen
  \bibfield  {author} {\bibinfo {author} {\bibfnamefont {A.~M.}\ \bibnamefont
  {Black-Schaffer}}\ and\ \bibinfo {author} {\bibfnamefont {J.}~\bibnamefont
  {Linder}},\ }\href {\doibase 10.1103/PhysRevB.84.180509} {\bibfield
  {journal} {\bibinfo  {journal} {Phys. Rev. B}\ }\textbf {\bibinfo {volume}
  {84}},\ \bibinfo {pages} {180509} (\bibinfo {year} {2011})}\BibitemShut
  {NoStop}%
\bibitem [{\citenamefont {Alicea}(2012)}]{alicea_new_2012}%
  \BibitemOpen
  \bibfield  {author} {\bibinfo {author} {\bibfnamefont {J.}~\bibnamefont
  {Alicea}},\ }\href {\doibase 10.1088/0034-4885/75/7/076501} {\bibfield
  {journal} {\bibinfo  {journal} {Rep. Prog. Phys.}\ }\textbf {\bibinfo
  {volume} {75}},\ \bibinfo {pages} {076501} (\bibinfo {year}
  {2012})}\BibitemShut {NoStop}%
\bibitem [{\citenamefont {Buzdin}(2008)}]{buzdin_direct_2008}%
  \BibitemOpen
  \bibfield  {author} {\bibinfo {author} {\bibfnamefont {A.}~\bibnamefont
  {Buzdin}},\ }\href {\doibase 10.1103/PhysRevLett.101.107005} {\bibfield
  {journal} {\bibinfo  {journal} {Phys. Rev. Lett.}\ }\textbf {\bibinfo
  {volume} {101}},\ \bibinfo {pages} {107005} (\bibinfo {year}
  {2008})}\BibitemShut {NoStop}%
\bibitem [{\citenamefont {Konschelle}\ and\ \citenamefont
  {Buzdin}(2009)}]{konschelle_magnetic_2009}%
  \BibitemOpen
  \bibfield  {author} {\bibinfo {author} {\bibfnamefont {F.}~\bibnamefont
  {Konschelle}}\ and\ \bibinfo {author} {\bibfnamefont {A.}~\bibnamefont
  {Buzdin}},\ }\href {\doibase 10.1103/PhysRevLett.102.017001} {\bibfield
  {journal} {\bibinfo  {journal} {Phys. Rev. Lett.}\ }\textbf {\bibinfo
  {volume} {102}},\ \bibinfo {pages} {017001} (\bibinfo {year}
  {2009})}\BibitemShut {NoStop}%
\bibitem [{\citenamefont {Ojanen}(2012)}]{ojanen_magnetoelectric_2012}%
  \BibitemOpen
  \bibfield  {author} {\bibinfo {author} {\bibfnamefont {T.}~\bibnamefont
  {Ojanen}},\ }\href {\doibase 10.1103/PhysRevLett.109.226804} {\bibfield
  {journal} {\bibinfo  {journal} {Phys. Rev. Lett.}\ }\textbf {\bibinfo
  {volume} {109}},\ \bibinfo {pages} {226804} (\bibinfo {year}
  {2012})}\BibitemShut {NoStop}%
\bibitem [{\citenamefont {Pershoguba}\ \emph {et~al.}(2015)\citenamefont
  {Pershoguba}, \citenamefont {Bj\"ornson}, \citenamefont {Black-Schaffer},\
  and\ \citenamefont {Balatsky}}]{pershoguba_currents_2015}%
  \BibitemOpen
  \bibfield  {author} {\bibinfo {author} {\bibfnamefont {S.~S.}\ \bibnamefont
  {Pershoguba}}, \bibinfo {author} {\bibfnamefont {K.}~\bibnamefont
  {Bj\"ornson}}, \bibinfo {author} {\bibfnamefont {A.~M.}\ \bibnamefont
  {Black-Schaffer}}, \ and\ \bibinfo {author} {\bibfnamefont {A.~V.}\
  \bibnamefont {Balatsky}},\ }\href {\doibase 10.1103/PhysRevLett.115.116602}
  {\bibfield  {journal} {\bibinfo  {journal} {Phys. Rev. Lett.}\ }\textbf
  {\bibinfo {volume} {115}},\ \bibinfo {pages} {116602} (\bibinfo {year}
  {2015})}\BibitemShut {NoStop}%
\bibitem [{\citenamefont {Chudnovsky}(2017)}]{chudnovsky_manipulating_2017}%
  \BibitemOpen
  \bibfield  {author} {\bibinfo {author} {\bibfnamefont {E.~M.}\ \bibnamefont
  {Chudnovsky}},\ }\href {\doibase 10.1103/PhysRevB.95.100503} {\bibfield
  {journal} {\bibinfo  {journal} {Phys. Rev. B}\ }\textbf {\bibinfo {volume}
  {95}},\ \bibinfo {pages} {100503} (\bibinfo {year} {2017})}\BibitemShut
  {NoStop}%
\bibitem [{\citenamefont {Konschelle}\ \emph {et~al.}(2015)\citenamefont
  {Konschelle}, \citenamefont {Tokatly},\ and\ \citenamefont
  {Bergeret}}]{konschelle_theory_2015}%
  \BibitemOpen
  \bibfield  {author} {\bibinfo {author} {\bibfnamefont {F.}~\bibnamefont
  {Konschelle}}, \bibinfo {author} {\bibfnamefont {I.~V.}\ \bibnamefont
  {Tokatly}}, \ and\ \bibinfo {author} {\bibfnamefont {F.~S.}\ \bibnamefont
  {Bergeret}},\ }\href {\doibase 10.1103/PhysRevB.92.125443} {\bibfield
  {journal} {\bibinfo  {journal} {Phys. Rev. B}\ }\textbf {\bibinfo {volume}
  {92}},\ \bibinfo {pages} {125443} (\bibinfo {year} {2015})}\BibitemShut
  {NoStop}%
\bibitem [{\citenamefont {Bergeret}\ and\ \citenamefont
  {Tokatly}(2015)}]{bergeret_theory_2015}%
  \BibitemOpen
  \bibfield  {author} {\bibinfo {author} {\bibfnamefont {F.~S.}\ \bibnamefont
  {Bergeret}}\ and\ \bibinfo {author} {\bibfnamefont {I.~V.}\ \bibnamefont
  {Tokatly}},\ }\href {\doibase 10.1209/0295-5075/110/57005} {\bibfield
  {journal} {\bibinfo  {journal} {EPL}\ }\textbf {\bibinfo {volume} {110}},\
  \bibinfo {pages} {57005} (\bibinfo {year} {2015})}\BibitemShut {NoStop}%
\bibitem [{\citenamefont {Fominov}\ \emph {et~al.}(2007)\citenamefont
  {Fominov}, \citenamefont {Volkov},\ and\ \citenamefont
  {Efetov}}]{fominov_josephson_2007}%
  \BibitemOpen
  \bibfield  {author} {\bibinfo {author} {\bibfnamefont {Y.~V.}\ \bibnamefont
  {Fominov}}, \bibinfo {author} {\bibfnamefont {A.~F.}\ \bibnamefont {Volkov}},
  \ and\ \bibinfo {author} {\bibfnamefont {K.~B.}\ \bibnamefont {Efetov}},\
  }\href {\doibase 10.1103/PhysRevB.75.104509} {\bibfield  {journal} {\bibinfo
  {journal} {Phys. Rev. B}\ }\textbf {\bibinfo {volume} {75}},\ \bibinfo
  {pages} {104509} (\bibinfo {year} {2007})}\BibitemShut {NoStop}%
\bibitem [{\citenamefont {Kulagina}\ and\ \citenamefont
  {Linder}(2014)}]{kulagina_spin_2014}%
  \BibitemOpen
  \bibfield  {author} {\bibinfo {author} {\bibfnamefont {I.}~\bibnamefont
  {Kulagina}}\ and\ \bibinfo {author} {\bibfnamefont {J.}~\bibnamefont
  {Linder}},\ }\href {\doibase 10.1103/PhysRevB.90.054504} {\bibfield
  {journal} {\bibinfo  {journal} {Phys. Rev. B}\ }\textbf {\bibinfo {volume}
  {90}},\ \bibinfo {pages} {054504} (\bibinfo {year} {2014})}\BibitemShut
  {NoStop}%
\bibitem [{\citenamefont {Konschelle}\ \emph {et~al.}(2016)\citenamefont
  {Konschelle}, \citenamefont {Tokatly},\ and\ \citenamefont
  {Bergeret}}]{konschelle_ballistic_2016}%
  \BibitemOpen
  \bibfield  {author} {\bibinfo {author} {\bibfnamefont {F.}~\bibnamefont
  {Konschelle}}, \bibinfo {author} {\bibfnamefont {I.~V.}\ \bibnamefont
  {Tokatly}}, \ and\ \bibinfo {author} {\bibfnamefont {F.~S.}\ \bibnamefont
  {Bergeret}},\ }\href {\doibase 10.1103/PhysRevB.94.014515} {\bibfield
  {journal} {\bibinfo  {journal} {Phys. Rev. B}\ }\textbf {\bibinfo {volume}
  {94}},\ \bibinfo {pages} {014515} (\bibinfo {year} {2016})}\BibitemShut
  {NoStop}%
\bibitem [{\citenamefont {Halterman}\ \emph {et~al.}(2008)\citenamefont
  {Halterman}, \citenamefont {Valls},\ and\ \citenamefont
  {Barsic}}]{halterman_induced_2008}%
  \BibitemOpen
  \bibfield  {author} {\bibinfo {author} {\bibfnamefont {K.}~\bibnamefont
  {Halterman}}, \bibinfo {author} {\bibfnamefont {O.~T.}\ \bibnamefont
  {Valls}}, \ and\ \bibinfo {author} {\bibfnamefont {P.~H.}\ \bibnamefont
  {Barsic}},\ }\href {\doibase 10.1103/PhysRevB.77.174511} {\bibfield
  {journal} {\bibinfo  {journal} {Phys. Rev. B}\ }\textbf {\bibinfo {volume}
  {77}},\ \bibinfo {pages} {174511} (\bibinfo {year} {2008})}\BibitemShut
  {NoStop}%
\bibitem [{\citenamefont {Braude}\ and\ \citenamefont
  {Nazarov}(2007)}]{braude_fully_2007}%
  \BibitemOpen
  \bibfield  {author} {\bibinfo {author} {\bibfnamefont {V.}~\bibnamefont
  {Braude}}\ and\ \bibinfo {author} {\bibfnamefont {Y.~V.}\ \bibnamefont
  {Nazarov}},\ }\href {\doibase 10.1103/PhysRevLett.98.077003} {\bibfield
  {journal} {\bibinfo  {journal} {Phys. Rev. Lett.}\ }\textbf {\bibinfo
  {volume} {98}},\ \bibinfo {pages} {077003} (\bibinfo {year}
  {2007})}\BibitemShut {NoStop}%
\bibitem [{\citenamefont {Mironov}\ and\ \citenamefont
  {Buzdin}(2017)}]{mironov_spontaneous_2017}%
  \BibitemOpen
  \bibfield  {author} {\bibinfo {author} {\bibfnamefont {S.}~\bibnamefont
  {Mironov}}\ and\ \bibinfo {author} {\bibfnamefont {A.}~\bibnamefont
  {Buzdin}},\ }\href {\doibase 10.1103/PhysRevLett.118.077001} {\bibfield
  {journal} {\bibinfo  {journal} {Phys. Rev. Lett.}\ }\textbf {\bibinfo
  {volume} {118}},\ \bibinfo {pages} {077001} (\bibinfo {year}
  {2017})}\BibitemShut {NoStop}%
\bibitem [{\citenamefont {Silaev}\ \emph {et~al.}(2017)\citenamefont {Silaev},
  \citenamefont {Tokatly},\ and\ \citenamefont
  {Bergeret}}]{silaev_anomalous_2017}%
  \BibitemOpen
  \bibfield  {author} {\bibinfo {author} {\bibfnamefont {M.~A.}\ \bibnamefont
  {Silaev}}, \bibinfo {author} {\bibfnamefont {I.~V.}\ \bibnamefont {Tokatly}},
  \ and\ \bibinfo {author} {\bibfnamefont {F.~S.}\ \bibnamefont {Bergeret}},\
  }\href {\doibase 10.1103/PhysRevB.95.184508} {\bibfield  {journal} {\bibinfo
  {journal} {Phys. Rev. B}\ }\textbf {\bibinfo {volume} {95}},\ \bibinfo
  {pages} {184508} (\bibinfo {year} {2017})}\BibitemShut {NoStop}%
\bibitem [{\citenamefont {Robinson}\ \emph {et~al.}(2010)\citenamefont
  {Robinson}, \citenamefont {Witt},\ and\ \citenamefont
  {Blamire}}]{robinson_controlled_2010}%
  \BibitemOpen
  \bibfield  {author} {\bibinfo {author} {\bibfnamefont {J.~W.~A.}\
  \bibnamefont {Robinson}}, \bibinfo {author} {\bibfnamefont {J.~D.~S.}\
  \bibnamefont {Witt}}, \ and\ \bibinfo {author} {\bibfnamefont {M.~G.}\
  \bibnamefont {Blamire}},\ }\href {\doibase 10.1126/science.1189246}
  {\bibfield  {journal} {\bibinfo  {journal} {Science}\ }\textbf {\bibinfo
  {volume} {329}},\ \bibinfo {pages} {59} (\bibinfo {year} {2010})}\BibitemShut
  {NoStop}%
\bibitem [{\citenamefont {Bergeret}\ and\ \citenamefont
  {Tokatly}(2014)}]{bergeret_spin-orbit_2014}%
  \BibitemOpen
  \bibfield  {author} {\bibinfo {author} {\bibfnamefont {F.~S.}\ \bibnamefont
  {Bergeret}}\ and\ \bibinfo {author} {\bibfnamefont {I.~V.}\ \bibnamefont
  {Tokatly}},\ }\href {\doibase 10.1103/PhysRevB.89.134517} {\bibfield
  {journal} {\bibinfo  {journal} {Phys. Rev. B}\ }\textbf {\bibinfo {volume}
  {89}},\ \bibinfo {pages} {134517} (\bibinfo {year} {2014})}\BibitemShut
  {NoStop}%
\bibitem [{\citenamefont {Mel’nikov}\ \emph {et~al.}(2012)\citenamefont
  {Mel’nikov}, \citenamefont {Samokhvalov}, \citenamefont {Kuznetsova},\ and\
  \citenamefont {Buzdin}}]{melnikov_interference_2012}%
  \BibitemOpen
  \bibfield  {author} {\bibinfo {author} {\bibfnamefont {A.~S.}\ \bibnamefont
  {Mel’nikov}}, \bibinfo {author} {\bibfnamefont {A.~V.}\ \bibnamefont
  {Samokhvalov}}, \bibinfo {author} {\bibfnamefont {S.~M.}\ \bibnamefont
  {Kuznetsova}}, \ and\ \bibinfo {author} {\bibfnamefont {A.~I.}\ \bibnamefont
  {Buzdin}},\ }\href {\doibase 10.1103/PhysRevLett.109.237006} {\bibfield
  {journal} {\bibinfo  {journal} {Phys. Rev. Lett.}\ }\textbf {\bibinfo
  {volume} {109}},\ \bibinfo {pages} {237006} (\bibinfo {year}
  {2012})}\BibitemShut {NoStop}%
\bibitem [{\citenamefont {Bogdanov}\ and\ \citenamefont
  {Yablonskii}(1989)}]{bogdanov_thermodynamicallystable_1989}%
  \BibitemOpen
  \bibfield  {author} {\bibinfo {author} {\bibfnamefont {A.~N.}\ \bibnamefont
  {Bogdanov}}\ and\ \bibinfo {author} {\bibfnamefont {D.~A.}\ \bibnamefont
  {Yablonskii}},\ }\href@noop {} {\bibfield  {journal} {\bibinfo  {journal}
  {Sov. Phys. JETP}\ }\textbf {\bibinfo {volume} {68}},\ \bibinfo {pages} {101}
  (\bibinfo {year} {1989})}\BibitemShut {NoStop}%
\bibitem [{\citenamefont {R\"o{\ss}ler}\ \emph {et~al.}(2011)\citenamefont
  {R\"o{\ss}ler}, \citenamefont {Leonov},\ and\ \citenamefont
  {Bogdanov}}]{rossler_chiral_2011}%
  \BibitemOpen
  \bibfield  {author} {\bibinfo {author} {\bibfnamefont {U.~K.}\ \bibnamefont
  {R\"o{\ss}ler}}, \bibinfo {author} {\bibfnamefont {A.~A.}\ \bibnamefont
  {Leonov}}, \ and\ \bibinfo {author} {\bibfnamefont {A.~N.}\ \bibnamefont
  {Bogdanov}},\ }\href {\doibase 10.1088/1742-6596/303/1/012105} {\bibfield
  {journal} {\bibinfo  {journal} {Journal of Physics: Conference Series}\
  }\textbf {\bibinfo {volume} {303}},\ \bibinfo {pages} {012105} (\bibinfo
  {year} {2011})}\BibitemShut {NoStop}%
\bibitem [{\citenamefont {Leonov}\ \emph {et~al.}(2016)\citenamefont {Leonov},
  \citenamefont {Monchesky}, \citenamefont {Romming}, \citenamefont {Kubetzka},
  \citenamefont {Bogdanov},\ and\ \citenamefont
  {Wiesendanger}}]{leonov_properties_2016}%
  \BibitemOpen
  \bibfield  {author} {\bibinfo {author} {\bibfnamefont {A.~O.}\ \bibnamefont
  {Leonov}}, \bibinfo {author} {\bibfnamefont {T.~L.}\ \bibnamefont
  {Monchesky}}, \bibinfo {author} {\bibfnamefont {N.}~\bibnamefont {Romming}},
  \bibinfo {author} {\bibfnamefont {A.}~\bibnamefont {Kubetzka}}, \bibinfo
  {author} {\bibfnamefont {A.~N.}\ \bibnamefont {Bogdanov}}, \ and\ \bibinfo
  {author} {\bibfnamefont {R.}~\bibnamefont {Wiesendanger}},\ }\href {\doibase
  10.1088/1367-2630/18/6/065003} {\bibfield  {journal} {\bibinfo  {journal}
  {New J. Phys.}\ }\textbf {\bibinfo {volume} {18}},\ \bibinfo {pages} {065003}
  (\bibinfo {year} {2016})}\BibitemShut {NoStop}%
\bibitem [{\citenamefont {Jonietz}\ \emph {et~al.}(2010)\citenamefont
  {Jonietz}, \citenamefont {M\"uhlbauer}, \citenamefont {Pfleiderer},
  \citenamefont {Neubauer}, \citenamefont {M\"unzer}, \citenamefont {Bauer},
  \citenamefont {Adams}, \citenamefont {Georgii}, \citenamefont {B\"oni},
  \citenamefont {Duine}, \citenamefont {Everschor}, \citenamefont {Garst},\
  and\ \citenamefont {Rosch}}]{jonietz_spin_2010}%
  \BibitemOpen
  \bibfield  {author} {\bibinfo {author} {\bibfnamefont {F.}~\bibnamefont
  {Jonietz}}, \bibinfo {author} {\bibfnamefont {S.}~\bibnamefont
  {M\"uhlbauer}}, \bibinfo {author} {\bibfnamefont {C.}~\bibnamefont
  {Pfleiderer}}, \bibinfo {author} {\bibfnamefont {A.}~\bibnamefont
  {Neubauer}}, \bibinfo {author} {\bibfnamefont {W.}~\bibnamefont {M\"unzer}},
  \bibinfo {author} {\bibfnamefont {A.}~\bibnamefont {Bauer}}, \bibinfo
  {author} {\bibfnamefont {T.}~\bibnamefont {Adams}}, \bibinfo {author}
  {\bibfnamefont {R.}~\bibnamefont {Georgii}}, \bibinfo {author} {\bibfnamefont
  {P.}~\bibnamefont {B\"oni}}, \bibinfo {author} {\bibfnamefont {R.~A.}\
  \bibnamefont {Duine}}, \bibinfo {author} {\bibfnamefont {K.}~\bibnamefont
  {Everschor}}, \bibinfo {author} {\bibfnamefont {M.}~\bibnamefont {Garst}}, \
  and\ \bibinfo {author} {\bibfnamefont {A.}~\bibnamefont {Rosch}},\
  }\href@noop {} {\bibfield  {journal} {\bibinfo  {journal} {Science}\ }\textbf
  {\bibinfo {volume} {330}} (\bibinfo {year} {2010})}\BibitemShut {NoStop}%
\bibitem [{\citenamefont {Yu}\ \emph {et~al.}(2012)\citenamefont {Yu},
  \citenamefont {Kanazawa}, \citenamefont {Zhang}, \citenamefont {Nagai},
  \citenamefont {Hara}, \citenamefont {Kimoto}, \citenamefont {Matsui},
  \citenamefont {Onose},\ and\ \citenamefont {Tokura}}]{yu_skyrmion_2012}%
  \BibitemOpen
  \bibfield  {author} {\bibinfo {author} {\bibfnamefont {X.}~\bibnamefont
  {Yu}}, \bibinfo {author} {\bibfnamefont {N.}~\bibnamefont {Kanazawa}},
  \bibinfo {author} {\bibfnamefont {W.}~\bibnamefont {Zhang}}, \bibinfo
  {author} {\bibfnamefont {T.}~\bibnamefont {Nagai}}, \bibinfo {author}
  {\bibfnamefont {T.}~\bibnamefont {Hara}}, \bibinfo {author} {\bibfnamefont
  {K.}~\bibnamefont {Kimoto}}, \bibinfo {author} {\bibfnamefont
  {Y.}~\bibnamefont {Matsui}}, \bibinfo {author} {\bibfnamefont
  {Y.}~\bibnamefont {Onose}}, \ and\ \bibinfo {author} {\bibfnamefont
  {Y.}~\bibnamefont {Tokura}},\ }\href
  {http://www.nature.com/articles/ncomms1990} {\bibfield  {journal} {\bibinfo
  {journal} {Nat. Commun.}\ }\textbf {\bibinfo {volume} {3}} (\bibinfo {year}
  {2012})}\BibitemShut {NoStop}%
\bibitem [{\citenamefont {Kiselev}\ \emph {et~al.}(2011)\citenamefont
  {Kiselev}, \citenamefont {Bogdanov}, \citenamefont {Sch\"afer},\ and\
  \citenamefont {R\"o{\ss}ler}}]{kiselev_chiral_2011}%
  \BibitemOpen
  \bibfield  {author} {\bibinfo {author} {\bibfnamefont {N.~S.}\ \bibnamefont
  {Kiselev}}, \bibinfo {author} {\bibfnamefont {A.~N.}\ \bibnamefont
  {Bogdanov}}, \bibinfo {author} {\bibfnamefont {R.}~\bibnamefont {Sch\"afer}},
  \ and\ \bibinfo {author} {\bibfnamefont {U.~K.}\ \bibnamefont
  {R\"o{\ss}ler}},\ }\href@noop {} {\bibfield  {journal} {\bibinfo  {journal}
  {J. Phys. D: Appl. Phys.}\ }\textbf {\bibinfo {volume} {44}} (\bibinfo {year}
  {2011})}\BibitemShut {NoStop}%
\bibitem [{\citenamefont {Iwasaki}\ \emph {et~al.}(2013)\citenamefont
  {Iwasaki}, \citenamefont {Mochizuki},\ and\ \citenamefont
  {Nagaosa}}]{iwasaki_current-induced_2013}%
  \BibitemOpen
  \bibfield  {author} {\bibinfo {author} {\bibfnamefont {J.}~\bibnamefont
  {Iwasaki}}, \bibinfo {author} {\bibfnamefont {M.}~\bibnamefont {Mochizuki}},
  \ and\ \bibinfo {author} {\bibfnamefont {N.}~\bibnamefont {Nagaosa}},\ }\href
  {\doibase 10.1038/nnano.2013.176} {\bibfield  {journal} {\bibinfo  {journal}
  {Nature Nanotech.}\ }\textbf {\bibinfo {volume} {8}},\ \bibinfo {pages} {742}
  (\bibinfo {year} {2013})}\BibitemShut {NoStop}%
\bibitem [{\citenamefont {Hrabec}\ \emph {et~al.}(2017)\citenamefont {Hrabec},
  \citenamefont {Sampaio}, \citenamefont {Belmeguenai}, \citenamefont {Gross},
  \citenamefont {Weil}, \citenamefont {Ch\'erif}, \citenamefont {Stashkevich},
  \citenamefont {Jacques}, \citenamefont {Thiaville},\ and\ \citenamefont
  {Rohart}}]{hrabec_current-induced_2017}%
  \BibitemOpen
  \bibfield  {author} {\bibinfo {author} {\bibfnamefont {A.}~\bibnamefont
  {Hrabec}}, \bibinfo {author} {\bibfnamefont {J.}~\bibnamefont {Sampaio}},
  \bibinfo {author} {\bibfnamefont {M.}~\bibnamefont {Belmeguenai}}, \bibinfo
  {author} {\bibfnamefont {I.}~\bibnamefont {Gross}}, \bibinfo {author}
  {\bibfnamefont {R.}~\bibnamefont {Weil}}, \bibinfo {author} {\bibfnamefont
  {S.~M.}\ \bibnamefont {Ch\'erif}}, \bibinfo {author} {\bibfnamefont
  {A.}~\bibnamefont {Stashkevich}}, \bibinfo {author} {\bibfnamefont
  {V.}~\bibnamefont {Jacques}}, \bibinfo {author} {\bibfnamefont
  {A.}~\bibnamefont {Thiaville}}, \ and\ \bibinfo {author} {\bibfnamefont
  {S.}~\bibnamefont {Rohart}},\ }\href {\doibase 10.1038/ncomms15765}
  {\bibfield  {journal} {\bibinfo  {journal} {Nat. Commun.}\ }\textbf {\bibinfo
  {volume} {8}},\ \bibinfo {pages} {15765} (\bibinfo {year}
  {2017})}\BibitemShut {NoStop}%
\bibitem [{\citenamefont {Fraerman}\ \emph {et~al.}(2005)\citenamefont
  {Fraerman}, \citenamefont {Karetnikova}, \citenamefont {Nefedov},
  \citenamefont {Shereshevskii},\ and\ \citenamefont
  {Silaev}}]{fraerman_magnetization_2005}%
  \BibitemOpen
  \bibfield  {author} {\bibinfo {author} {\bibfnamefont {A.~A.}\ \bibnamefont
  {Fraerman}}, \bibinfo {author} {\bibfnamefont {I.~R.}\ \bibnamefont
  {Karetnikova}}, \bibinfo {author} {\bibfnamefont {I.~M.}\ \bibnamefont
  {Nefedov}}, \bibinfo {author} {\bibfnamefont {I.~A.}\ \bibnamefont
  {Shereshevskii}}, \ and\ \bibinfo {author} {\bibfnamefont {M.~A.}\
  \bibnamefont {Silaev}},\ }\href {\doibase 10.1103/PhysRevB.71.094416}
  {\bibfield  {journal} {\bibinfo  {journal} {Phys. Rev. B}\ }\textbf {\bibinfo
  {volume} {71}},\ \bibinfo {pages} {094416} (\bibinfo {year}
  {2005})}\BibitemShut {NoStop}%
\bibitem [{\citenamefont {Vadimov}\ \emph {et~al.}(2018)\citenamefont
  {Vadimov}, \citenamefont {Sapozhnikov},\ and\ \citenamefont
  {Mel'nikov}}]{vadimov_magnetic_2018}%
  \BibitemOpen
  \bibfield  {author} {\bibinfo {author} {\bibfnamefont {V.~L.}\ \bibnamefont
  {Vadimov}}, \bibinfo {author} {\bibfnamefont {M.~V.}\ \bibnamefont
  {Sapozhnikov}}, \ and\ \bibinfo {author} {\bibfnamefont {A.~S.}\ \bibnamefont
  {Mel'nikov}},\ }\href {\doibase 10.1063/1.5037934} {\bibfield  {journal}
  {\bibinfo  {journal} {Appl. Phys. Lett.}\ }\textbf {\bibinfo {volume}
  {113}},\ \bibinfo {pages} {032402} (\bibinfo {year} {2018})}\BibitemShut
  {NoStop}%
\bibitem [{\citenamefont {Rabinovich}\ \emph {et~al.}(2018)\citenamefont
  {Rabinovich}, \citenamefont {Bobkova}, \citenamefont {Bobkov},\ and\
  \citenamefont {Silaev}}]{rabinovich_chirality_2018}%
  \BibitemOpen
  \bibfield  {author} {\bibinfo {author} {\bibfnamefont {D.~S.}\ \bibnamefont
  {Rabinovich}}, \bibinfo {author} {\bibfnamefont {I.~V.}\ \bibnamefont
  {Bobkova}}, \bibinfo {author} {\bibfnamefont {A.~M.}\ \bibnamefont {Bobkov}},
  \ and\ \bibinfo {author} {\bibfnamefont {M.~A.}\ \bibnamefont {Silaev}},\
  }\href {\doibase 10.1103/PhysRevB.98.184511} {\bibfield  {journal} {\bibinfo
  {journal} {Phys. Rev. B}\ }\textbf {\bibinfo {volume} {98}},\ \bibinfo
  {pages} {184511} (\bibinfo {year} {2018})}\BibitemShut {NoStop}%
\bibitem [{\citenamefont {Yang}\ \emph {et~al.}(2016)\citenamefont {Yang},
  \citenamefont {Stano}, \citenamefont {Klinovaja},\ and\ \citenamefont
  {Loss}}]{yang_majorana_2016}%
  \BibitemOpen
  \bibfield  {author} {\bibinfo {author} {\bibfnamefont {G.}~\bibnamefont
  {Yang}}, \bibinfo {author} {\bibfnamefont {P.}~\bibnamefont {Stano}},
  \bibinfo {author} {\bibfnamefont {J.}~\bibnamefont {Klinovaja}}, \ and\
  \bibinfo {author} {\bibfnamefont {D.}~\bibnamefont {Loss}},\ }\href {\doibase
  10.1103/PhysRevB.93.224505} {\bibfield  {journal} {\bibinfo  {journal} {Phys.
  Rev. B}\ }\textbf {\bibinfo {volume} {93}},\ \bibinfo {pages} {224505}
  (\bibinfo {year} {2016})}\BibitemShut {NoStop}%
\bibitem [{\citenamefont {G\"ung\"ord\"u}\ \emph {et~al.}(2018)\citenamefont
  {G\"ung\"ord\"u}, \citenamefont {Sandhoefner},\ and\ \citenamefont
  {Kovalev}}]{gungordu_stabilization_2018}%
  \BibitemOpen
  \bibfield  {author} {\bibinfo {author} {\bibfnamefont {U.}~\bibnamefont
  {G\"ung\"ord\"u}}, \bibinfo {author} {\bibfnamefont {S.}~\bibnamefont
  {Sandhoefner}}, \ and\ \bibinfo {author} {\bibfnamefont {A.~A.}\ \bibnamefont
  {Kovalev}},\ }\href {\doibase 10.1103/PhysRevB.97.115136} {\bibfield
  {journal} {\bibinfo  {journal} {Phys. Rev. B}\ }\textbf {\bibinfo {volume}
  {97}},\ \bibinfo {pages} {115136} (\bibinfo {year} {2018})}\BibitemShut
  {NoStop}%
\bibitem [{\citenamefont {Takashima}\ and\ \citenamefont
  {Fujimoto}(2016)}]{takashima_supercurrent-induced_2016}%
  \BibitemOpen
  \bibfield  {author} {\bibinfo {author} {\bibfnamefont {R.}~\bibnamefont
  {Takashima}}\ and\ \bibinfo {author} {\bibfnamefont {S.}~\bibnamefont
  {Fujimoto}},\ }\href {\doibase 10.1103/PhysRevB.94.235117} {\bibfield
  {journal} {\bibinfo  {journal} {Phys. Rev. B}\ }\textbf {\bibinfo {volume}
  {94}},\ \bibinfo {pages} {235117} (\bibinfo {year} {2016})}\BibitemShut
  {NoStop}%
\bibitem [{\citenamefont {Pershoguba}\ \emph {et~al.}(2016)\citenamefont
  {Pershoguba}, \citenamefont {Nakosai},\ and\ \citenamefont
  {Balatsky}}]{pershoguba_skyrmion-induced_2016}%
  \BibitemOpen
  \bibfield  {author} {\bibinfo {author} {\bibfnamefont {S.~S.}\ \bibnamefont
  {Pershoguba}}, \bibinfo {author} {\bibfnamefont {S.}~\bibnamefont {Nakosai}},
  \ and\ \bibinfo {author} {\bibfnamefont {A.~V.}\ \bibnamefont {Balatsky}},\
  }\href {\doibase 10.1103/PhysRevB.94.064513} {\bibfield  {journal} {\bibinfo
  {journal} {Phys. Rev. B}\ }\textbf {\bibinfo {volume} {94}},\ \bibinfo
  {pages} {064513} (\bibinfo {year} {2016})}\BibitemShut {NoStop}%
\bibitem [{\citenamefont {Hals}\ \emph {et~al.}(2016)\citenamefont {Hals},
  \citenamefont {Schecter},\ and\ \citenamefont
  {Rudner}}]{hals_composite_2016}%
  \BibitemOpen
  \bibfield  {author} {\bibinfo {author} {\bibfnamefont {K.~M.}\ \bibnamefont
  {Hals}}, \bibinfo {author} {\bibfnamefont {M.}~\bibnamefont {Schecter}}, \
  and\ \bibinfo {author} {\bibfnamefont {M.~S.}\ \bibnamefont {Rudner}},\
  }\href {\doibase 10.1103/PhysRevLett.117.017001} {\bibfield  {journal}
  {\bibinfo  {journal} {Phys. Rev. Lett.}\ }\textbf {\bibinfo {volume} {117}},\
  \bibinfo {pages} {017001} (\bibinfo {year} {2016})}\BibitemShut {NoStop}%
\bibitem [{\citenamefont {Nagaosa}\ and\ \citenamefont
  {Tokura}(2013)}]{nagaosa_topological_2013}%
  \BibitemOpen
  \bibfield  {author} {\bibinfo {author} {\bibfnamefont {N.}~\bibnamefont
  {Nagaosa}}\ and\ \bibinfo {author} {\bibfnamefont {Y.}~\bibnamefont
  {Tokura}},\ }\href {\doibase 10.1038/nnano.2013.243} {\bibfield  {journal}
  {\bibinfo  {journal} {Nature Nanotech.}\ }\textbf {\bibinfo {volume} {8}},\
  \bibinfo {pages} {899} (\bibinfo {year} {2013})}\BibitemShut {NoStop}%
\bibitem [{\citenamefont {Lyuksyutov}\ and\ \citenamefont
  {Pokrovsky}(2005)}]{lyuksyutov_ferromagnet_2005}%
  \BibitemOpen
  \bibfield  {author} {\bibinfo {author} {\bibfnamefont {I.~F.}\ \bibnamefont
  {Lyuksyutov}}\ and\ \bibinfo {author} {\bibfnamefont {V.~L.}\ \bibnamefont
  {Pokrovsky}},\ }\href {\doibase 10.1080/00018730500057536} {\bibfield
  {journal} {\bibinfo  {journal} {Adv. Phys.}\ }\textbf {\bibinfo {volume}
  {54}},\ \bibinfo {pages} {67} (\bibinfo {year} {2005})}\BibitemShut {NoStop}%
\bibitem [{\citenamefont {Gennes}(1966)}]{gennes_superconductivity_1966}%
  \BibitemOpen
  \bibfield  {author} {\bibinfo {author} {\bibfnamefont {P.~G.~D.}\
  \bibnamefont {Gennes}},\ }\href@noop {} {\emph {\bibinfo {title}
  {Superconductivity of {Metals} and {Alloys}}}}\ (\bibinfo  {publisher}
  {Benjamin},\ \bibinfo {year} {1966})\BibitemShut {NoStop}%
\bibitem [{\citenamefont {Edelstein}(1995)}]{edelstein_magnetoelectric_1995}%
  \BibitemOpen
  \bibfield  {author} {\bibinfo {author} {\bibfnamefont {V.~M.}\ \bibnamefont
  {Edelstein}},\ }\href {\doibase 10.1103/PhysRevLett.75.2004} {\bibfield
  {journal} {\bibinfo  {journal} {Phys. Rev. Lett.}\ }\textbf {\bibinfo
  {volume} {75}},\ \bibinfo {pages} {2004} (\bibinfo {year}
  {1995})}\BibitemShut {NoStop}%
\bibitem [{\citenamefont {Edelstein}(1996)}]{edelstein_ginzburg_1996}%
  \BibitemOpen
  \bibfield  {author} {\bibinfo {author} {\bibfnamefont {V.~M.}\ \bibnamefont
  {Edelstein}},\ }\href {\doibase 10.1088/0953-8984/8/3/012} {\bibfield
  {journal} {\bibinfo  {journal} {J. Phys. Condens. Matter}\ }\textbf {\bibinfo
  {volume} {8}},\ \bibinfo {pages} {339} (\bibinfo {year} {1996})}\BibitemShut
  {NoStop}%
\bibitem [{\citenamefont {Samokhin}(2004)}]{samokhin_magnetic_2004}%
  \BibitemOpen
  \bibfield  {author} {\bibinfo {author} {\bibfnamefont {K.~V.}\ \bibnamefont
  {Samokhin}},\ }\href {\doibase 10.1103/PhysRevB.70.104521} {\bibfield
  {journal} {\bibinfo  {journal} {Phys. Rev. B}\ }\textbf {\bibinfo {volume}
  {70}},\ \bibinfo {pages} {104521} (\bibinfo {year} {2004})}\BibitemShut
  {NoStop}%
\bibitem [{\citenamefont {Kaur}\ \emph {et~al.}(2005)\citenamefont {Kaur},
  \citenamefont {Agterberg},\ and\ \citenamefont
  {Sigrist}}]{kaur_helical_2005}%
  \BibitemOpen
  \bibfield  {author} {\bibinfo {author} {\bibfnamefont {R.~P.}\ \bibnamefont
  {Kaur}}, \bibinfo {author} {\bibfnamefont {D.~F.}\ \bibnamefont {Agterberg}},
  \ and\ \bibinfo {author} {\bibfnamefont {M.}~\bibnamefont {Sigrist}},\ }\href
  {\doibase 10.1103/PhysRevLett.94.137002} {\bibfield  {journal} {\bibinfo
  {journal} {Phys. Rev. Lett.}\ }\textbf {\bibinfo {volume} {94}},\ \bibinfo
  {pages} {137002} (\bibinfo {year} {2005})}\BibitemShut {NoStop}%
\bibitem [{\citenamefont {Hrabec}\ \emph {et~al.}(2014)\citenamefont {Hrabec},
  \citenamefont {Porter}, \citenamefont {Wells}, \citenamefont {Benitez},
  \citenamefont {Burnell}, \citenamefont {McVitie}, \citenamefont {McGrouther},
  \citenamefont {Moore},\ and\ \citenamefont
  {Marrows}}]{hrabec_measuring_2014}%
  \BibitemOpen
  \bibfield  {author} {\bibinfo {author} {\bibfnamefont {A.}~\bibnamefont
  {Hrabec}}, \bibinfo {author} {\bibfnamefont {N.~A.}\ \bibnamefont {Porter}},
  \bibinfo {author} {\bibfnamefont {A.}~\bibnamefont {Wells}}, \bibinfo
  {author} {\bibfnamefont {M.~J.}\ \bibnamefont {Benitez}}, \bibinfo {author}
  {\bibfnamefont {G.}~\bibnamefont {Burnell}}, \bibinfo {author} {\bibfnamefont
  {S.}~\bibnamefont {McVitie}}, \bibinfo {author} {\bibfnamefont
  {D.}~\bibnamefont {McGrouther}}, \bibinfo {author} {\bibfnamefont {T.~A.}\
  \bibnamefont {Moore}}, \ and\ \bibinfo {author} {\bibfnamefont {C.~H.}\
  \bibnamefont {Marrows}},\ }\href {\doibase 10.1103/PhysRevB.90.020402}
  {\bibfield  {journal} {\bibinfo  {journal} {Phys. Rev. B}\ }\textbf {\bibinfo
  {volume} {90}},\ \bibinfo {pages} {020402} (\bibinfo {year}
  {2014})}\BibitemShut {NoStop}%
\bibitem [{\citenamefont {Dimitrova}\ and\ \citenamefont
  {Feigel'man}(2007)}]{dimitrova_theory_2007}%
  \BibitemOpen
  \bibfield  {author} {\bibinfo {author} {\bibfnamefont {O.}~\bibnamefont
  {Dimitrova}}\ and\ \bibinfo {author} {\bibfnamefont {M.~V.}\ \bibnamefont
  {Feigel'man}},\ }\href {\doibase 10.1103/PhysRevB.76.014522} {\bibfield
  {journal} {\bibinfo  {journal} {Phys. Rev. B}\ }\textbf {\bibinfo {volume}
  {76}},\ \bibinfo {pages} {014522} (\bibinfo {year} {2007})}\BibitemShut
  {NoStop}%
\end{thebibliography}%

\end{document}